\documentstyle[11pt,newpasp,twoside]{article}
\def\edcomment#1{\iffalse\marginpar{\raggedright\sl#1\/}\else\relax\fi}
\reversemarginpar

\begin{document}
\title{T Dwarf Discoveries by 2MASS}
 \author{Adam J. Burgasser}
\affil{California Institute of Technology, MSC 103-33, Pasadena, CA 91125}

\begin{abstract}
Over a dozen T dwarfs, brown dwarfs that exhibit methane absorption
features at 1.6 and
2.2 $\micron$, have been discovered by the 2MASS survey.  We discuss
how the search for these objects has been made, point out some of 
the limitations of
using near-infrared data to find the warmest T dwarfs, and 
highlight a few of the more
interesting T dwarfs that have been identified in 2MASS data. 
\end{abstract}

\section{Introduction}

T dwarfs are brown dwarfs that exhibit CH$_4$ absorption features
at K-band
(Kirkpatrick et al.\ 1999).  These objects have effective temperatures
T$_{eff}$ $\la$ 1300 - 1500 K (Fegley \& Lodders 1996),
and are thus the coolest brown dwarfs so far detected.  
Until recently, only one object of this 
type was known, Gl 229B (Nakajima et al.\ 1995), a low-luminosity companion
with T$_{eff}$ $\sim$ 1000 K and substellar mass M $\sim$ 40 
M$_{Jup}$\footnote{M$_{Jup}$ $\equiv$ 1 Jupiter mass = 
1.9 $\times$ 10$^{30}$ g $\approx$ 
0.1 M$_{\sun}$.} (Allard et al.\ 1996; Marley et al.\ 1996).
The first field T dwarf, SDSS 1624+00, was identified by 
Strauss et al.\ (1999) in the 
Sloan Digital Sky Survey (SDSS; York et al.\ 2000); additional 
discoveries were soon made
using the Two Micron All Sky Survey
(2MASS; Burgasser et al.\ 1999, 2000a, 2000c, 2001b), SDSS (Tsvetanov et al.\ 2000;
Leggett et al.\ 2000), and the New Technology Telescope Deep Field 
(Cuby et al.\ 1999).
The detection of so many cool brown dwarfs (23 at the time of this writing)
emphasizes how ubiquitous these objects are likely to be, despite their only
recent identification; Reid et al.\ (1999)
estimate that there may be up to twice as many brown dwarfs in the Galaxy as 
stars. 

In this article, we focus on 2MASS T dwarf discoveries, of which 16 
have so far been made.  In $\S$2 we discuss the search process used to
identify T dwarfs in 2MASS near-infrared (NIR) data, and some of the limitations 
encountered in identifying warm T dwarfs (i.e., L/T ``transition objects'').
In $\S$$\S$3-5 we summarize three discoveries: the coolest known brown dwarf
Gl 570D,
the H$\alpha$ emitter 2MASS 1237+65, and the
brightest T dwarf so far identified, 2MASS 0559-14.
We summarize future progress in $\S$6. 

\section{The 2MASS T Dwarf Search}

CH$_4$, H$_2$O and H$_2$
collision-induced (CIA) absorption force the spectral energy distributions
of T dwarfs to peak in the NIR at J-band (1.25 
$\micron$).  As a case in point, Gl 229B has an
absolute J-band magnitude M$_J$ = 15.4 (1.1 mJy at 1.25 $\micron$), while 
M$_{r}$ $>$ 23.3 ($<$ 0.0014 mJy at 0.7 $\micron$), M$_{i}$ = 21.2 (0.0084 mJy
at 0.85 $\micron$), M$_{K}$ = 15.6 (0.38 mJy at 2.2 $\micron$) and
M$_{L'}$ = 13.4 (0.13 mJy at 3.7 $\micron$;
Matthews et al.\ 1996; Leggett et al.\ 1999).  
This is in contrast to the 
F$_{\nu}$ peak wavelength 
of 5 $\micron$ for a 1000 K blackbody.  
NIR surveys, such as
2MASS (Skrutskie et al.\ 1997), are thus
especially tuned to detect the maximum flux from these otherwise
dim objects.  2MASS is currently
completing a whole-sky survey to SNR = 10 limits of J = 15.8, H = 15.1,
and K$_s$ = 14.3; at these magnitudes, it is capable of detecting 
objects like Gl 229B out to 
distances of 12, 8.3, and 5.8 pc, respectively, although fainter sources are
certainly attainable.  

\begin{figure}
\plotfiddle{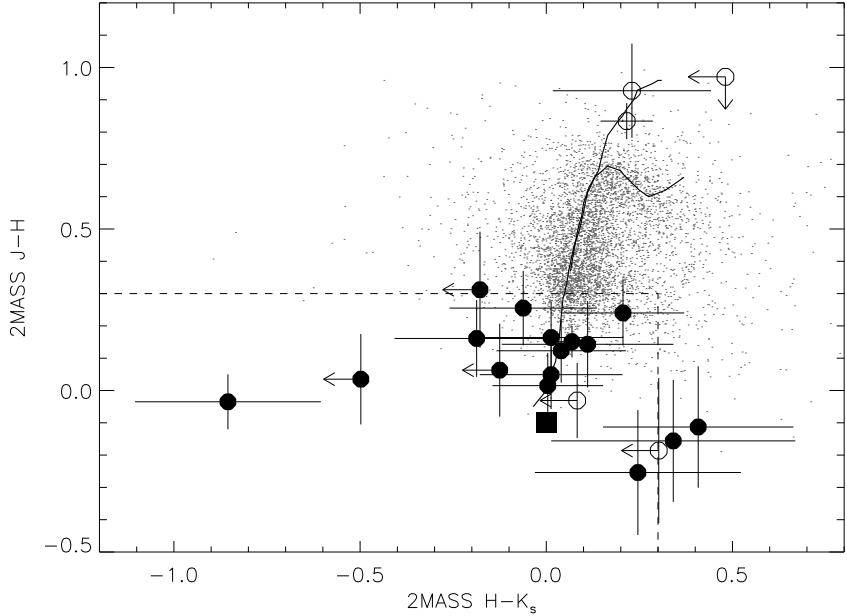}{3in}{90}{50}{50}{190}{-40}
\caption{NIR color-color diagram of 2MASS point sources (points), T
dwarf discoveries by 2MASS (filled circles) and SDSS (open circles), and
Gl 229B (square). 
Arrows indicate upper limits.  The Bessell \& Brett
(1988) dwarf and giant tracks are overlaid (solid lines), 
as are the color constraints
made for the June 1999 sample (dashed lines).}
\end{figure}

Figure 1 shows a NIR color-color plot of 
a sample of point sources obtained from the 2MASS Second Incremental Data
Release\footnote{http://www.ipac.caltech.edu/2mass/releases/second/index.html}.
The positions of currently known T dwarfs are indicated,
along with the Bessell \& Brett (1988) dwarf and giant 
tracks.   
Note that 2MASS T dwarf discoveries (filled circles)
begin at the blue end of the dwarf track, 
while warm T dwarfs identified by SDSS (open circles at upper right) 
lie along an extension of the 
giant track redward of J-H $\ga$ 0.8.  Between these lies the NIR main
sequence, and the majority of point sources detected by 2MASS.
The contamination of
these background sources forces us to constrain our search to J-H and H-K$_s$
$\leq$ 0.3 (dashed lines).  Based on 2MASS completeness limits, 
we set a J-band magnitude limit of 16 and require J- and H-band detections,
although no constraint is made at K$_s$-band.  The faintness of 
Gl 229B in the optical (R - J $\sim$ 9; Golimowski et al.\ 1998)
allows us to rule 
out objects with associated optical counterparts 
(USNO-A2.0 catalog; Monet et al.\ 1998); we also eliminate
catalogued minor planets.  Finally, candidates are constrained to
Galactic latitudes 
${\mid}b^{II}{\mid}$ $\geq$ 15$\deg$ in order to avoid source confusion.  

In June 1999, we extracted 48,339 candidates from
18,360 sq.\ deg.\ (44.5\% of the sky) of 2MASS point source data.  
Optical images of each candidate field,
obtained from the CADC DSS 
server\footnote{http://cadcwww.dao.nrc.ca/cadcbin/getdss},
were examined to rule out faint counterparts
or close doubles.  Second-epoch
NIR imaging of remaining candidates was then done 
to eliminate uncatalogued minor planets, which have similar colors as T
dwarfs.
Finally, NIR spectroscopy 
confirmed T dwarf status via 
identification of the 1.6 and 2.2 $\micron$ CH$_4$ bands.  

Figure 2 shows NIR spectral data 
for a sample of 2MASS T dwarfs,  
obtained at the Keck 10m using the Near-Infrared
Camera (NIRC; Matthews \& Soifer 1994),
along with data for Gl 229B 
(Oppenheimer et al.\ 1998).  
Molecular absorption bands
of CH$_4$ and H$_2$O are clearly evident, while suppression of flux at K-band
is also due to H$_2$ CIA.  The slight bend in the
J-band peak at 1.25 $\micron$ is due to an unresolved K I doublet, 
which is clearly
evident in higher resolution data (Strauss et al.\ 1999).
One striking
feature of these spectra is that they are {\em not} identical.
The prominent 1.6 $\micron$ CH$_4$ band shows a gradual deepening from
2MASS 0559-14 to Gl 570D, as do combined CH$_4$ and H$_2$O absorption bands
at 1.15 and 1.4 $\micron$.  At the same time, the 
1.25 $\micron$ K I feature disappears, 
while the K-band peak changes from a sharp
bandhead at 2.2 $\micron$ to a more gradual drop-off longward of 
2.1 $\micron$.
We believe these spectral morphology changes are 
primarily due to temperature differences
between the objects.
Deeper molecular bands initially result from increasing CH$_4$ and H$_2$O 
number densities in the photosphere via the reaction
CO + 3H$_2$ $\rightarrow$ CH$_4$ + H$_2$O.
At lower temperatures, the reduction of higher energy ``wing'' transitions
increases the contrast between the band cores and adjacent continuum.  
K I doublet lines at 1.2432 and 1.2522 $\micron$, caused by the 
higher order 4p $^2$P$_0$ - 5s $^2$S transition, weaken with decreasing temperatures,
and perhaps additionally through conversion of atomic potassium into KCl 
(Lodders 1999).  Increased H$_2$ opacity toward lower temperatures
(Lenzuni, Chernoff, \& Salpeter 1991) results in increased absorption beyond
2.1 $\micron$. 
 
\begin{figure}
\plotfiddle{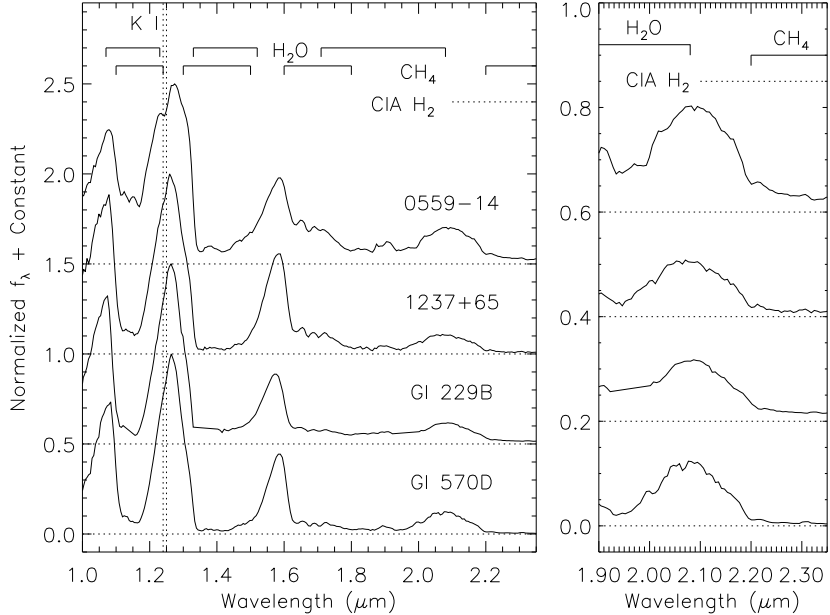}{3in}{90}{50}{50}{190}{-40}
\caption{(Left)
NIRC spectra of 2MASS T dwarfs and Gl 229B (Oppenheimer et al.\ 1998).
Major absorption features of CH$_4$, H$_2$O, H$_2$, and K I are indicated.
(Right) Close-up view of the K-band peaks.}
\end{figure}
 
In all, 13 T dwarfs have now
been identified in the June 1999 sample, 
which has been nearly completely followed up (Burgasser et al.\ 2001b).
Table 1 shows the number of detections, area surveyed, 
and estimated space
densities of this sample and those of Burgasser et al.\ (1999), 
Strauss et al.\ (1999) and Tsvetanov et al.\ (2000), 
and Leggett et al.\ (2000).
Note that the SDSS space densities are generally higher than those from 
2MASS searches.  While this discrepancy
may be due to small number statistics in the SDSS results,
the color constraints imposed by background sources in
2MASS data certainly eliminates some T dwarfs.    
SDSS i*-z* colors delineate cool brown dwarfs
due to their extreme redness from 0.8--1 $\micron$ (Fan et al.\ 2000;
Figure 3 below), quite unlike main sequence stars.
As such, SDSS should be able
to produce a more complete sample of T dwarfs by including the warm 
L/T ``transition'' objects,
while 2MASS likely samples a larger volume
due to greater detectability at J-band.  

\begin{table}
\caption{T dwarf Detections in Various Surveys.}
\begin{tabular}{r|cccc}
\tableline
\tableline
Sample\tablenotemark{a} & June 1999 & B99 & ST00 & Le00 \\
\tableline
No.\ Detections & 13 & 4 & 2 & 3 \\
Area Searched (sq.\ deg.) & 18360 & 1784 & 130 & 225 \\
Est. Space Density (pc$^{-3}$) & 0.003 & 0.010 & 0.045 & 0.0095 \\
\tableline
\end{tabular}
\tablenotetext{a}{B99 = Burgasser et al.\ (1999); 
ST00 = Strauss et al.\ (1999) and Tsvetanov et al.\ (2000); 
Le00 = Leggett et al.\ (2000).}
\end{table}

\section{Gliese 570D}

Gl 570D (Burgasser et al.\ 2000a) was identified 
in the June 1999 sample as a field T dwarf candidate, but is 
located only 258$\farcs$3 $\pm$
0$\farcs$4 from Gl 570ABC.  This triple system consists of a
K4V primary separated by $\sim$ 25$\arcsec$ from a close 
M1.5V-M3V binary.  
NIR spectral data for 
Gl 570D are shown in Figure 2.
Deep CH$_4$
and H$_2$O absorption bands, as well as the absence of the 1.25 $\micron$ K I 
feature, indicate that it is very cool object.

Companionship was
verified via common proper motion by 
a second 2MASS observation of the field
14 months after the initial catalog imaging.  
The proper motion of the system (2$\farcs$012 $\pm$ 0$\farcs$002 yr$^{-1}$;
Perryman et al.\ 1997) was clearly resolved, given the 
0$\farcs$3 astrometric accuracy of 2MASS point source
data\footnote{R.\ M.\ Cutri et al.\ 1999, Explanatory Supplement of the
2MASS Incremental Data Release 
(http://www.ipac.caltech.edu/2mass/releases/second/doc/explsup.html).}.
Based on the distance to Gl 570A (5.91 $\pm$ 0.06 pc; Perryman et al.\ 1997),
Gl 570D has M$_J$ = 16.47 $\pm$
0.07, nearly a full magnitude dimmer than Gl 229B 
(M$_J$ = 15.51 $\pm$ 0.09; Leggett et al.\ 1999).
Estimates based on empirical bolometric corrections, as well results from
evolutionary models (Burrows et al.\ 1997) yield T$_{eff}$ = 750 $\pm$ 50 K, 
($\sim$ 200 K cooler than Gl 229B) and M = 50 $\pm$ 20 M$_{Jup}$.  
Gl 570D is the coolest brown dwarf
so far detected; but, because the Gl 570 system is rather
old ($\tau$ $\sim$ 2-10 Gyr), it may actually be more massive than 
Gl 229B.  Companion brown dwarfs such as these, including recent L dwarf
companion discoveries (Kirkpatrick et al.\ 2000b; Wilson et al.\ 2000), will help
tie spectral types to absolute magnitudes, as well as provide age
information -- valuable
empirical constraints for evolutionary models.

\section{2MASS 1237+65}

2MASS 1237+65 (Burgasser et al.\ 1999) is a faint (J = 16.03 $\pm$ 0.09) T dwarf
undetected at K$_s$-band by 2MASS.  Optical spectra of this and two other 
T dwarfs (SDSS 1624+00 and SDSS 1346-00; Strauss et al.\ 1999; 
Tsvetanov et al.\ 2000) were obtained using the Keck 10m 
Low Resolution Imaging Spectrograph (LRIS; Oke et al.\ 1995),
in the wavelength range 6300 to 10100 {\AA} at 9 {\AA} resolution.  
Figure 3 shows the 
reduced spectra, with prominent features of Cs I, FeH, CH$_4$, and H$_2$O
indicated.  The broad absorption responsible for the rapid ramp-up from 8000 to
10000 {\AA} is likely due to the 
pressure-broadened K I doublet at 7665 and 7699
{\AA} (Burrows et al.\ 2000; Liebert et al.\ 2000).  

\begin{figure}
\plotfiddle{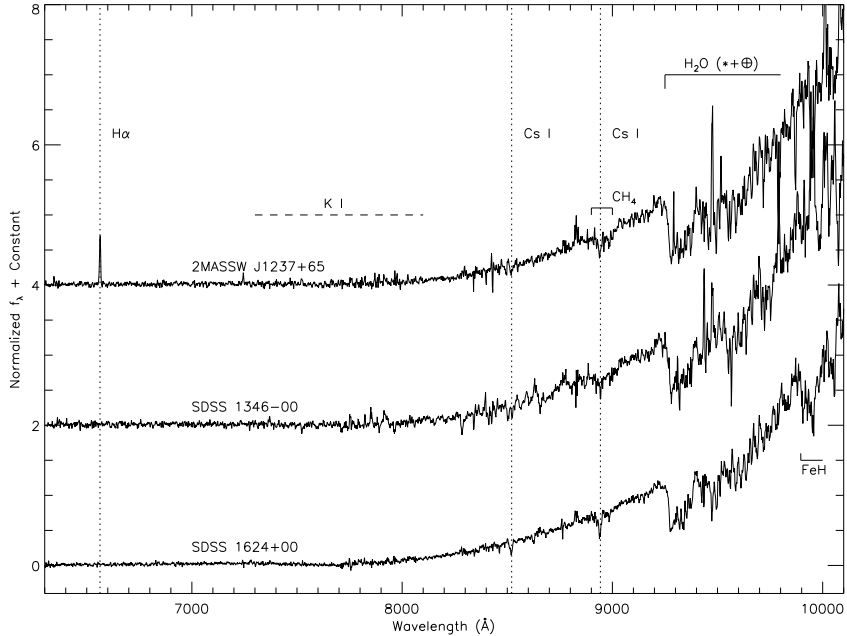}{3in}{90}{50}{50}{190}{-40}
\caption{Optical spectra of 2MASS 1237+65, SDSS 1624+00, and SDSS 1346-00
(From Burgasser et al.\ 2000c).}
\end{figure}

Also indicated in Figure 3 is H$\alpha$ emission at 6563 {\AA}, clearly
seen in 2MASS 1237+65.  This emission line was observed to persist over three days,
and subsequent observations have confirmed its presence over timescales
exceeding  
6 months.  While H$\alpha$ emission is prevalent in late M dwarfs and early
L dwarfs, Kirkpatrick et al.\ (2000a) and Gizis et al.\ (2000) have
shown that the incidence and strength 
of emission drops off precipitously beyond spectral type 
M7V, with no objects later then L5V showing H$\alpha$ in emission.  Observations
of other T dwarfs (Burgasser 2001a) show a similar lack of
activity as measured by the H$\alpha$ line.  By estimating the bolometric flux
of 2MASS 1237+65,
we measure log($f_{H{\alpha}}/f_{bol}$) = log($L_{H{\alpha}}/L_{bol}$) = $-$4.2,
consistent with the activity level of late dMe field stars (Hawley et al.\ 1996).
Given these clues, 2MASS 1237+65 must have a unique activity mechanism in
order to sustain its relatively high level of emission.  We have
proposed that 2MASS 1237+65
may be a close, interacting binary, in which a lower mass secondary
(and thereby {\em larger}
brown dwarf, as $dlnR/dlnM$ = -1/3 for degenerate dwarfs) is losing mass to the
primary via Roche lobe overflow,  
causing shock heating in the latter's atmosphere.  In order 
to sustain mass-loss,
the physical separation of the primary and secondary must be
on the order of 10 Jupiter radii; thus, photometric monitoring will
likely provide the best test for this scenario.  
Regardless, the activity in 2MASS 1237+65
remains an intriguing puzzle.

\section{2MASS 0559-14}

The majority of T dwarfs identified by 2MASS and SDSS have J-band
magnitudes between 15 and 16, placing them between 6 pc (for the coolest
objects such as Gl 570D) and $\sim$ 15 pc, depending
on temperature and duplicity.  Nevertheless, we have identified one object, 
2MASS 0559-14 (Burgasser et al.\ 2000c), 
which is significantly brighter, with J = 13.83 $\pm$ 0.03.
This is over one magnitude brighter
than any other 2MASS T dwarf so far discovered, 
and nearly 0.4 mag brighter than Gl 229B at J-band.
In addition, 2MASS 0559-14 is the reddest T dwarf identified  
in the June 1999 sample, 
with J-K$_s$ = 0.22 $\pm$ 0.06.  
A rough estimate based on the absolute J-band magnitudes of the L8V
Gl 584C (15.0; Kirkpatrick et al.\ 2000a) and Gl 229B (15.5; Leggett et al.\ 1999),
places 2MASS 0559-14 at about 5 pc (if it is single)\footnote{This 
argument assumes that M$_J$
monotonically decreases from L8V through type T.  This relation may not hold
if CH$_4$ and H$_2$O bands strengthen significantly over a small T$_{eff}$
range, possibly redistributing flux into the J-band window.}. 

NIR spectral data for 2MASS 0559-14 is shown in Figure 2.
The most striking feature about this object is its significant deviation in
spectral morphology from the other T dwarfs shown.  
While cool brown dwarfs such
as Gl 229B and Gl 570D show deep H$_2$O and CH$_4$ absorption bands at 1.15,
1.4, 1.6--2.0, and 2.2 $\micron$, 2MASS 0559-14 has significantly shallower
absorption troughs, most notably at the 1.6 $\micron$ CH$_4$ band.  We
argue that these features, along with detectable 0.9896 $\micron$
FeH (which weakens toward the
late L dwarfs) and strong 1.2 $\micron$ K I, 
suggest that 2MASS 0559-14 is significantly
warmer than other 2MASS discoveries.  Leggett et al.\ (2000) have recently 
identified
three SDSS T dwarfs which have even 
shallower CH$_4$ absorption at H- and K-bands, as well as 
a residual CO bandhead at 2.3 $\micron$.  
These objects are probably warmer still, right
at the transition between the latest L dwarfs and the onset of CH$_4$ absorption in
the NIR\footnote{Noll et al.\ (2000) have recently detected 
the 3.3 $\micron$
fundamental band of CH$_4$ in objects as early as L5V, indicating that
some conversion of CO to CH$_4$ may occur prior to
the formation of the 1.6 and 2.2 $\micron$ overtone bands.}.

\section{Future Progress}

In just over one year, an entire population of cool T dwarfs has
been discovered.  A great deal remains to be learned about these objects,
including absolute magnitudes, space motion, 
duplicity, variability,
activity, space density, and ultimately their contribution to the
stellar/substellar mass function and the mass budget of the Galaxy.
While over a dozen T dwarfs have been identified in
2MASS data, more than 
one-half of the sky and a great deal of color space remains 
to be investigated.  In addition, as SDSS begins routine survey operations,
the opportunity to cross-correlate 2MASS and SDSS data will allow more
sophisticated and hence more complete searches for brown dwarfs in the field.
Given the estimates shown in Table 1, it is likely that we will find 
anywhere from 2 to 20 T dwarfs within 5 pc of the Sun, significantly 
influencing the nearby star sample.
Finally, it is now possible to develop a new classification for T dwarfs, in
analogy to the L dwarf spectral class derived by Kirkpatrick et al.\ (1999).
Ironically, the future looks quite bright for brown dwarf research.


\begin{references}
\reference Allard, F., Hauschildt, P.\ H., Baraffe, I., \& Chabrier, G.\
1996, \apj, 465, L123
\reference Bessell, M.\ S., \& Brett, J.\ M.\ 1988, \pasp, 100, 1134
\reference Burgasser, A.\ J., et al.\ 1999, \apj, 522, L65
\reference Burgasser, A.\ J., et al.\ 2000a, \apj, 531, L57
\reference Burgasser, A.\ J., Kirkpatrick, J.\ D., Reid, I.\ N., 
Liebert, J., Gizis, J.\ E., \& Brown, M.\ E.\ 2000b, \aj, 120, 473
\reference Burgasser, A.\ J., et al.\ 2000c, \aj, 120, 1100
\reference Burgasser, A.\ J. 2001a, these proceedings
\reference Burgasser, A.\ J., et al.\ 2001b, \apj, in prep.\
\reference Burrows, A., et al.\ 1997, \apj, 491, 856
\reference Burrows, A., Marley, M.\ S., \& Sharp, C.\ M.\ 2000, \apj, 531, 438
\reference Cuby, J.\ G., Saracco, P., Moorwood, A.\ F.\ M., D'Odorico, S.,
Lidman, C., Comer{\'{o}}n, F., \& Spyromilio, J.\ 1999, \aap, 349, L41
\reference Fan, X., et al.\ 2000, \aj, 119, 928
\reference Fegley, B., \& Lodders, K.\ 1996, \apj, 472, L37
\reference Gizis, J.\ E., Monet, D.\ G., Reid, I.\ N., Kirkpatrick, J.\ D.,
Liebert, J., \& Williams, R.\ J.\ 2000, \aj, 120, 1085
\reference Golimowski, D.\ A., Burrows, C.\ J., Kulkarni, S.\ R., 
Oppenheimer, B.\ R., \& Brukardt, R.\ A.\ 1998, \aj, 115, 2579
\reference Hawley, S.\ L., Gizis, J.\ E., \& Reid, I.\ N.\ 1996, \aj, 112, 2799
\reference Kirkpatrick, J.\ D., et al.\ 1999, \apj, 519, 802
\reference Kirkpatrick, J.\ D., Reid, I.\ N., Liebert, J., Gizis, J.\ E.,
Burgasser, A.\ J., Monet, D.\ G., Dahn, C.\ C., Nelson, B., \& Williams, R.\ J.\
2000a, \aj, 120, 447 
\reference Kirkpatrick, J.\ D., et al.\ 2000b, \aj, in prep.\ 
\reference Leggett, S.\ K., Toomey, D.\ W., Geballe, T.\ R., \& Brown,
R.\ H.\ 1999, \apj, 517, L139 
\reference Leggett, S.\ K., et al.\ 2000, \apj, 536, L35
\reference Lenzuni, P., Chernoff, D.\ F., \& Salpeter, E.\ E.\ 1991, \apjs, 
76, 759
\reference Liebert, J., Reid, I.\ N., Burrows, A., Burgasser, A.\ J., 
Kirkpatrick, J.\ D., \& Gizis, J.\ E.\ 2000, \apj, 533, L155
\reference Lodders, K.\ 1999, \apj, 519, 793
\reference Marley, M., Saumon, D., Guillot, T., Freedman, R.\ S.,
Hubbard, W.\ B., Burrows, A., \& Lunine, J.\ I.\ 1996, Science, 272, 1919
\reference Matthews, K., \& Soifer, B.\ T.\ 1994, in Infrared Astronomy
with Arrays: The Next Generation, ed.\ I.\ McLean (Dordrecht: Kluwer), 239
\reference Matthews, K., Nakajima, T., Kulkarni, S.\ R., \& Oppenheimer,
B.\ R.\ 1996, \aj, 112, 1678
\reference Monet, D.\ G., et al.\ 1998, USNO-A2.0 Catalog (Flagstaff: USNO)
\reference Nakajima, T., Oppenheimer, B.\ R., Kulkarni, S.\ R.,
Golimowski, D.\ A., Matthews, K., \& Durrance, S.\ T.\ 1995, Nature, 378, 463
\reference Noll, K.\ S., Geballe, T.\ R., Leggett, S.\ K., \& Marley, M.\ S.\
2000, \apj, in press
\reference Oke, J.\ B., et al.\ 1995, \pasp, 107, 375
\reference Oppenheimer, B.\ R., Kulkarni, S.\ R., Matthews, K., \& van Kerkwijk,
M.\ H.\ 1998, \apj, 502, 932
\reference Perryman, M.\ A.\ C., et al.\ 1997, \aap, 323, L49
\reference Reid, I.\ N., et al.\ 1999, \apj, 521, 631
\reference Skrutskie, M.\ F., et al.\ 1997, in The Impact of Large-Scale
Near-IR Sky Surveys, ed.\ F.\ Garzon (Dordrecht: Kluwer), 25
\reference Strauss, M.\ A., et al.\ 1999, \apj, 522, L61
\reference Tsvetanov, Z.\ I., et al.\ 2000, \apj, 531, L61
\reference Wilson, J.\ C., et al.\ 2000, \aj, in prep.\
\reference York, D.\ G., et al.\ 2000, \aj, 120, 1579
\end{references}
\end{document}